\documentclass[%
 reprint,
 amsmath,amssymb,
 aps,
 prl,
 nobibnotes
]{revtex4-1}

\usepackage{graphicx}% Include figure files
\usepackage{dcolumn}% Align table columns on decimal point
\usepackage{bm}% bold math
\usepackage{xcolor}

\begin{document}

\preprint{APS/123-QED}

\title{Phase-sensitive symmetry breaking in bidirectionally pumped Kerr microresonators}% Force line breaks with \\

\author{Elena A. Anashkina and Alexey V. Andrianov}
\altaffiliation{The authors contributed equally to this work}
\affiliation{A.V. Gaponov-Grekhov Institute of Applied Physics of the Russian Academy of Sciences Nizhny Novgorod, 603950, Russia} 
%

% \email{email@gmail.com}

\date{\today}% It is always \today, today,

\begin{abstract}

We demonstrate theoretically and experimentally that linear coupling due  to the backscattering between two counterpropagating modes in a Kerr microresonator leads to extreme sensitivity of the intensity asymmetry of nonlinear light states to the relative phase between the bidirectional pumps of equal power. In the absence of linear coupling, the relative pump phase does not affect counterpropagating intraresonator intensities, and two asymmetric states arise due to spontaneous symmetry breaking. Contrariwise, in the presence of weak linear intermode coupling, the asymmetry of the states is deterministically controlled via changes in phase (for all phases except 0 and $\pi$). Spontaneous symmetry breaking at zero phase is suppressed when the linear intermode coupling is increased, while for the $\pi$ phase it can be enhanced, so that the overall threshold for the spontaneous symmetry breaking can be significantly lower for nonzero linear coupling. 
These results are important for fundamental understanding of the processes in Kerr resonators and other systems with Kerr-like nonlinearities and linear intermode coupling and have high prospects for the development of photonic devices such as ultrasensitive sensors.

\end{abstract}

\maketitle
\begin{figure}
\includegraphics[width=0.95\columnwidth]{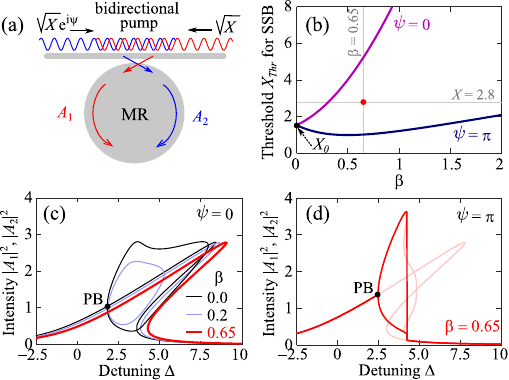}
\caption{\label{fig:L1} (a) Schematic diagram of the considered system with bidirectional two-mode propagation. (b) Threshold pump power $X_{Thr}$ for SSB via a pitchfork bifurcation (PB) for $\psi = 0$ and $\psi = \pi$ as a function of $\beta$. $X_{Thr}(0) = X_0 = 8/(3\sqrt{3})$. The red dot corresponds to parameters $X = 2.8$ and $\beta = 0.65$ (predominantly studied in this Letter). (c) Stationary theoretical solutions of Eqs.~(\ref{eq:Eq1}) for pump power in each direction $X = 2.8$, relative phase between pumps $\psi = 0$, and different intermode coupling coefficient $\beta$. (d) Stationary theoretical solutions (light red curve) and numerically simulated realisation of $|A_1|^2$ and $|A_2|^2$ for increasing detuning (red curves) for $X = 2.8$, $\beta = 0.65$ and $\psi = \pi$.}
\end{figure}

Spontaneous symmetry breaking (SSB) is a fundamentally important concept known in many areas of physics: particle physics, condensed matter physics, and optics, to name a few \cite{IntroSSB,Electroweak,BEC, Malomed, BoseHubb}. In photonics, there has been an increasing interest in studying  symmetry breaking, chiral and nonreciprocal light propagation in nonlinear optical macro- and microresonators \cite{Malomed,Pascal2017, Pascal, Erkintalo, PRL2017, Pascal2024}. A model of a resonant system with two originally degenerate modes that can interact via third-order nonlinearity and may be pumped individually with two signals has become a widely adopted theoretical and experimental platform for exploring symmetry breaking and accompanying phenomena in different physical realizations. In particular, Kerr microresonators with counterpropagating modes and bidirectional pump (Fig. \ref{fig:L1}(a)) are convenient for experimental studies. Besides, they demonstrate an enormous potential for numerous applications such as non-magnetic optical isolators and circulators  \cite{Optica2018}, ultra-high-sensitivity sensors \cite{indexsensor} and gyroscopes \cite{gyro1, gyro2}, photonic flip-flops, and many others.

When counterpropagating modes interact only incoherently due to Kerr nonlinearity, the relative phase between the pumps of equal intensities do not affect the dynamics of the intraresonator intensities. It is known that such a system demonstrates SSB: a symmetric state (with equal intensities) becomes unstable after a certain power threshold, and a pair of asymmetric states appears \cite{Kaplan}. Which one of these states is realized is determined by residual asymmetries in the system or random fluctuations. More recent experimental and theoretical works have shown that the dynamics of such a system can be very rich; multistable states including the ones with broken symmetry can occur; at high pump powers, regular and chaotic oscillations and complex bifurcations may be observed \cite{Pascal, PRE}. 

However, coherent linear coupling between counterpropagating modes is often present in real systems, resulting from backscattering (e.g., Rayleigh scattering) \cite{Pryam}. Scattering in resonators plays a key role in ultrasensitive detection and characterisation of nanoobjects \cite{nanoparticle,molecscatter}, as well as in fundamental studies, such as coupling between an atom and a microresonator \cite{atom} and cavity quantum electrodynamics \cite{QED}. 

Linear coupling due to backscattering is reciprocal and symmetric in magnitude (i.e., the power scattering coefficient from mode $A_1$ to $A_2$ is equal to the scattering coefficient from mode $A_2$ to $A_1$). Nevertheless, due to the coherent nature of the backscattering signal, the phases of counterpropagating waves become interlinked. This leads to overall sensitivity to the phase difference between the pumps, which becomes especially important for nonlinear phenomena, such as symmetry breaking. Surprisingly, the phase sensitivity of such systems has not been previously studied, to the best of our knowledge.

Here, we demonstrate that in a Kerr microresonator even small coherent coupling between counterpropagating modes (leading to resonance splitting smaller than their linewidth) drastically alters the nonlinear system dynamics under the same pump powers, so that the formation of asymmetric states becomes sensitive to the phase $\psi$ between the bidirectional pumps. The difference between the intensities of the counterporpagating modes (asymmetry) is deterministic for all phases except for two critical values 0 and $\pi$, at which the asymmetry changes its sign and can have a discontinuity. These discontinuities correspond to SSB, similar to that observed in a system with uncoupled modes. Symmetry breaking near a zero phase is suppressed when the coupling is increased, while for a phase near $\pi$ it can be enhanced, so that the overall threshold for SSB is lower for a nonzero linear coupling. In addition, the presence of discontinuities results in extreme phase sensitivity near critical values, which can be useful for sensing applications.  

We consider two counterpropagating modes $A_1$ and $A_2$ in a microresonator with instantaneous Kerr nonlinearity. They are excited by bidirectional pumps with the same powers and the relative phase $\psi$ between them  (Fig.~\ref{fig:L1}(a)). Linear coupling between $A_1$ and $A_2$ with the coefficient $\beta$ is added to the model \cite{Pryam, Lobanov}. Our system can be described by the following dynamic equations in dimensionless form \cite{Pascal, PRE, Lobanov}: 
\begin{subequations}
	\label{eq:Eq1}
\begin{eqnarray}
\frac{dA_1}{dt} = (-1-i\Delta+i|A_1|^2+2i|A_2|^2)A_1+
\nonumber \\
i\beta_{12} A_2 + \sqrt{X},\\
\frac{dA_2}{dt} = (-1-i\Delta+2i|A_1|^2+i|A_2|^2)A_2+
\nonumber \\
i\beta_{21} A_1 + \sqrt{X}e^{i\psi},
\end{eqnarray}
\end{subequations}
where $t$ is the slow time, $\Delta$ is the pump frequency detuning, and $X$ is the power of each pump. The terms $-A_{1,2}$ describe linear losses, the terms $i|A_{1,2}|^2 A_{1,2}$ and $2i|A_{1,2}|^2 A_{2,1}$ describe Kerr self-phase modulation and cross-phase modulation, respectively. In most cases coupling coefficients can be represented as $\beta_{21}^*=\beta_{12}=|\beta|\exp(i \chi)$, where $\chi$ is related to the phase of the backscattered wave. Making the transformation $A_2 \rightarrow A_2 \exp(i \chi)$ we encapsulate it in the pump phase $\psi \rightarrow \psi+\chi$.
In what follows, we consider $\beta$ to be real and positive. 
The case $\beta = 0$ has been well studied theoretically \cite{Pascal, PRE} and used to explain experimental results \cite{Pascal2017, Pascal}.

We find all stationary states of the system of equations~\ref{eq:Eq1}(a,b) by setting $d/dt = 0$ and solving numerically nonlinear algebraic equations for $|A_1|^2$ and $|A_2|^2$. However, not all stationary states are stable and accessible for certain combinations of parameters and initial conditions. So, numerical simulations of the dynamical system with very slowly varying detuning are also implemented to emulate experiments. Moreover, numerical modeling reveals oscillating regimes.  
% "However... " - can be deleted

First of all, let us recall the behavior of the system at $\beta = 0$ \cite{Kaplan, PRE, Pascal2017}. For this case,  $|A_1|^2$, $|A_2|^2$ are independent of the phase. So, in the previous works, the zero phase was often assumed without discussion. The SSB via pitchfork bifurcation occurs for pump powers higher than the threshold  $X_0 = 8/(3\sqrt{3}) \approx 1.54$ \cite{Kaplan, PRE, Pascal2017} (this value is shown by the black dot in Fig.~\ref{fig:L1}(b) at $\beta = 0$). The phenomenon originates from the unequal Kerr effects for the wave itself and the counterpropagating wave: cross-phase modulation is twice as strong as self-phase modulation.  %"The phenomenon..." - can be deleted
As an illustrative example we plotted stationary states of $|A_1|^2$ and $|A_2|^2$ as functions of detuning at moderate pump power $X=2.8$ (Fig.~\ref{fig:L1}(c), black curves). In a certain range of detunings ($\sim\!1.8<\Delta <~\sim\!7.9$), the symmetric solution becomes unstable, and there is a pair of stable asymmetric branches.

Now let us turn to $\beta \neq 0$. The special cases $\psi=0, \psi=\pi$ deserve separate attention. Although in both cases the system is still symmetric with respect to the exchange $|A_1|^2 \leftrightarrow |A_2|^2$, the difference in its behavior is striking.  
We found threshold powers $X_{Thr}$ vs $\beta$ for SSB and plotted the curves  in Fig.~\ref{fig:L1}(b). 
In our example $X=2.8$, an increase in $\beta$ at $\psi=0$ leads to narrowing of the range of detunings $\Delta$ corresponding to SSB, and the symmetry is eventually fully restored for $\beta > \sim 0.31$ for all detunings (Fig.~\ref{fig:L1}(c)). 
For $\psi=0$, $X_{Thr}$ grows fast starting from its absolute minimum $X_0$ at $\beta=0$. Contrariwise, for $\psi=\pi$,   $X_{Thr}$ starts from the same $X_0$ value, but first drops to 1 at $\beta=0.5$ and then grows slowly.

We will use for illustration the value $\beta=0.65$, which is not a very strong coupling corresponding to just a slight broadening of the linear resonance. This regime has been realized in our experiment. The pump power of $X=2.8$ is below the threshold at $\psi=0$, but well above the threshold at $\psi=\pi$ (Fig.~\ref{fig:L1}(b, red dot)). 
The corresponding stationary states of $|A_1|^2$, $|A_2|^2$ at $\psi=\pi$ with several symmetry broken branches are presented in Fig.~\ref{fig:L1}(d) with light red curves. 
To reveal the  branches that are realized in a common experimental situation, where the detuning is scanned from negative (pump frequency above the cold resonance) to positive (pump frequency below the cold resonance) values, we independently integrated the dynamic equations~(\ref{eq:Eq1})(a,b). We gradually increased $\Delta$ for each subsequent run and set $A_1$ and $A_2$ found at the previous detuning value as the initial conditions for the next detuning value. The corresponding steady states $|A_1|^2$ and $|A_2|^2$ found by numerical integration of the dynamic system are shown by red curves in Fig.~\ref{fig:L1}(d) (phase difference between the modes is discussed in supplementary materials). The pitchfork bifurcation occurs at $\Delta \approx  2.4$; it is randomly determined which wave has higher intensity. After the branches with broken symmetry end, the solution jumps to the lowest symmetric branch. Other solutions are not realized for such scanning of the detuning.

The characteristic features of the symmetry breaking near $\psi=0$ (the formation of a symmetry broken state and return
to the same branch of the stable symmetric state as the detuning increases) are close to those described in the literature for $\beta=0$ 
(Fig.~\ref{fig:L1}(c)). However, for nonzero $\beta$ symmetry breaking near $\pi$ appears at lower powers and it is more prominent in a wide range of pump powers and coupling coefficients (Fig.~\ref{fig:L1}(b)). Its characteristic features include abrupt termination of the asymmetric branches followed by the jump directly to the low-power symmetric state and overall higher asymmetry (see Fig.~\ref{fig:L1}(d)). 
 
\begin{figure}
\includegraphics[width=0.95\columnwidth]{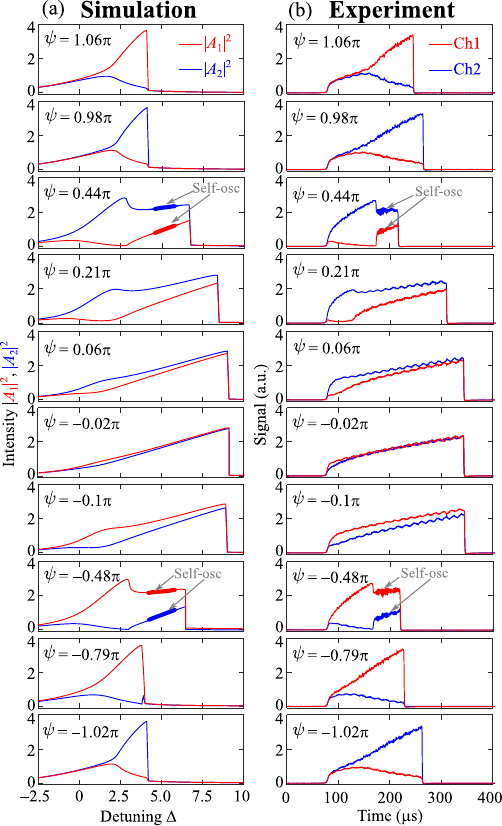}
\caption{\label{fig:L4new} (a) Intensities $|A_1|^2$ and $|A_2|^2$ for different $\psi$ numerically simulated for increasing detuning for $X = 2.8$ and $\beta = 0.65$. Thick lines for $\psi = 0.44\pi$ and $\psi = -0.48\pi$ correspond to self-oscillation regimes (without specifying values of their amplitude). (b) Experimental signals measured by oscilloscope in channel 1 (red curves) and channel 2 (blue curves) while scanning the pump frequency for different relative phases $\psi$ between pumps indicated at each panel. }
\end{figure}

Further, let us discuss the system behavior at $X=2.8$, $\beta =0.65$ when $\psi\neq0,\pi$. Here Eqs.~\ref{eq:Eq1}(a,b) lose their symmetry with respect to $|A_1|^2 \leftrightarrow |A_2|^2$. The stable solutions obtained by numerical integration are presented in Fig.~\ref{fig:L4new}(a). For small $|\psi|$, the solutions for $|A_1|^2$ and $|A_2|^2$ are only slightly deformed relative to the symmetric solutions at $\psi = 0$. The intensity in the second mode $|A_2|^2$ is slightly higher than in the first mode $|A_1|^2$ at $\psi>0$ (and vice versa at $\psi<0$). As the relative phase increases, the asymmetry becomes more significant and strongly depends on $\Delta$. The structure of the stationary solutions becomes very complex (refer to the supplementary materials for more details). However, modeling of a dynamic system shows that with slow frequency scanning (by increasing $\Delta$), a relatively simple scenario is realized. The intensity $|A_2|^2$ (blue curves in Fig.~\ref{fig:L4new}(a)) is deterministically higher than the intensity $|A_1|^2$ (red curves) for $0 < \psi < \pi$ and vice versa for $-\pi < \psi < 0$. The corresponding experimental results are presented in Fig.~\ref{fig:L4new}(b).

\begin{figure}
\includegraphics[width=0.95\columnwidth]{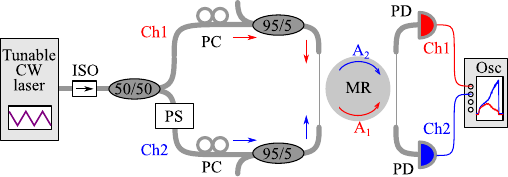}
\caption{\label{fig:L3} (a) Simplified experimental scheme. ISO – Faraday isolator; PS – phase shifter; PC – polarization controller; MR – microresonator; PD – photodiode; Osc – oscilloscope; Ch1 – channel 1; Ch2 – channel 2.}
\end{figure}

To confirm our theoretical results we conducted an experiment according to a simplified scheme shown in Fig.~\ref{fig:L3}(a). A silica microsphere with a diameter of 120 $\mu$m was bidirectionally pumped by a tunable laser through a fiber taper. To measure intraresonator fields we used a second taper in a ``drop-port" configuration and two identical photodetectors. 
The phase difference between the pumps introduced by a piezoelectric phase shifter was incremented from $-1.5\pi$ to $1.5 \pi$ in 40 small steps. At each step the laser frequency was scanned up and then down in the range from about $-10$ to 10 GHz. 
All experiments were done for a resonance with the loaded Q-factor of about $10^8$, and Rayleigh scattering coefficient $\beta = 0.65$ estimated from linear measurements (see supplementary for details).

Nonlinear responses at a pump power of about 5 mW in each channel as functions of time for different phases $\psi$ are shown in Fig.\ref{fig:L4new}(b)).The difference between the frequencies of the laser and the “cold” resonance did not exactly correspond to detuning $\Delta$, since the resonances experienced a dynamic thermal shift during scanning \cite{Therm}. The thermal shift was exactly the same for both waves, so it did not affect the symmetry, and $\Delta$ was a monotonic function of time. Keeping this in mind, a very good agreement was achieved between the experiment and simulations regarding the asymmetry of light states for the corresponding phases (compare Fig.\ref{fig:L4new} (a) and (b)). At a phase close to zero ($\psi = -0.02$), the curves for both waves are very close. For $0 < \psi < \pi$, the nonlinear resonance curve for the wave in channel 2 is higher than for the wave in channel 1, and for $-\pi < \psi < 0$, the resonance curve for the wave in channel 1 is higher. The frames with $\psi=-0.48 \pi$ and $0.44 \pi$ show the jump to the middle branch after passing the maximum of the topmost branch, and also the development of self-oscillations (see supplementary materials). When passing through $\pm\pi$, the curves change places. 

\begin{figure}
\includegraphics[width=0.95\columnwidth]{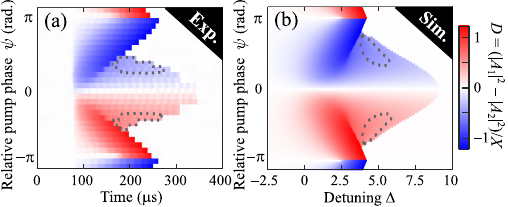}
\caption{\label{fig:L5} (a) Normalized experimental intensity difference of counterpropagating waves as a function of time and phase $\psi$. (b) Normalized intensity difference ($D = (|A_1|^2 - |A_2|^2)/X$) as a function of  $\Delta$ and $\psi$ numerically simulated for increasing detuning for $X = 2.8$ and $\beta = 0.65$ (for oscillating modes averaged values are taken). Areas inside dotted curves in (a) and (b) correspond to self-oscillation regimes.  }
\end{figure}

Good visualization of all measured and numerically simulated  frames is provided by plotting  the intensity difference of counterpropagating waves normalized to the pump intensity ($D = (|A_1|^2 - |A_2|^2)/X$) as a function of detuning and relative pump phase (Fig.~\ref{fig:L5}). The system evolves smoothly through the symmetric state ($|A_1|^2=|A_1|^2$) as the phase $\psi$ passes 0, develops the asymmetry as the phase goes away from 0 and experiences an abrupt flip of asymmetry (swap of intensities $|A_1|^2 \leftrightarrow |A_2|^2$, i.e. $D \rightarrow -D$) at $\psi=\pm \pi$.  The whole graph is symmetric about the lines $\psi=0$ and $\psi=\pm\pi$, which indicates a good balance of pump powers and absence of other asymmetries in the system except for the pump phase difference. In the modeling, if for some parameters an oscillation regime was realized, we took the time-averaged (stationary) values $|A_1|^2$, $|A_2|^2$ to calculate $D$.

\begin{figure}
\includegraphics[width=0.95\columnwidth]{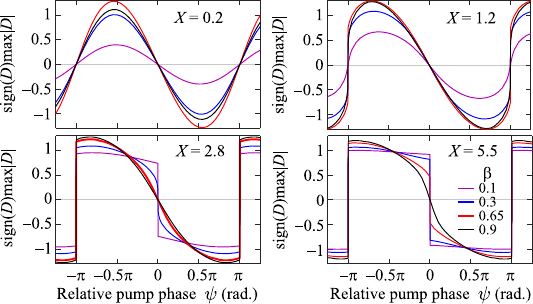}
\caption{\label{fig:xbeta}  Numerically simulated normalized difference of intensities of counterpropagating waves maximized over detuning  $\operatorname{sign}(D)\max|D|$ as a function of phase $\psi$ for different $\beta$. $D=(|A_1|^2-|A_2|^2)/X$. Each panel corresponds to different $X$.} 
\end{figure}

Finally, we discuss a more general picture of the asymmetry of light states in a wide range of $\beta$ and $X$. In numerical modeling we found the maximum of $|D|$ over all detunings and plotted $\operatorname{sign}(D)\max|D|$
as a function of $\psi$ for different coupling coefficients and pump powers (Fig.~\ref{fig:xbeta}). For nonzero $\beta$ the asymmetry  was observed even for small pump power ($X=0.2$ corresponds to almost linear regime). The dependence on the phase was smooth, and the sign of asymmetry was fully deterministic.  For a moderate power, a discontinuity was formed at $\psi=\pi$ for a certain $\beta$, which corresponds to nondeterministic SSB at this point. For an even higher power, another discontinuity (hence, nondeterministic symmetry breaking) could also be observed at $\psi=0$; however, it was easily suppressed by increasing coupling. As pump power grew, the region of maximum asymmetry shifted from $\psi=\pi/2$ towards $\psi=\pi$.

Even with weak linear intermode coupling, the system loses its symmetry for all phases except 0 and $\pi$; therefore, the asymmetric states become deterministic. Arguably, in systems with extremely weak coupling, when it becomes merely a perturbation of the same order as other imperfections and/or disturbances, the symmetry breaking is no longer fully deterministic, but random and triggered by fluctuations other than the pump phase difference. This brings us back to the regimes well described in earlier works that did not take into account intermodal coupling.

An important feature is extreme sensitivity of the system to the phase, which is indicated by jumps in the asymmetry near 0 and $\pi$. Formally, the sensitivity diverges at these points. The system acts a phase comparator that converts a small phase difference between the pumps into a binary output. As was discussed earlier, the phase $\psi$ is a combination of the phase difference between the pumps and the phase of the mode coupling coefficient. When coupling occurs due to a localized scatterer such as a nanoprobe \cite{Nanoscatterer}, a nanoparticle \cite{nanoparticle}, or even individual molecules \cite{molecscatter}, the phase depends on the position of the scatterer on the microresonator circumference with respect to the pump taper coupling point \cite{Nanoscatterer}. Therefore, the enhanced phase sensitivity can help to track and/or adjust the position of the scatterer with precision much higher than the wavelength. 

To conclude,  in Kerr microresonators with counterpropagating modes, linear mode coupling results in a nontrivial dependence of the system dynamics on the phase difference between the pumps and introduces new regimes that can not be viewed as simple perturbations of known dynamics in an uncoupled system. This must be considered when planning, conducting and interpreting experiments involving symmetric excitation of resonant systems with Kerr-like nonlinearity and two degenerate modes, even if linear coupling between modes is not expected to be strong.
The extreme phase sensitivity of symmetry breaking in nonlinear resonant systems induced by linear mode coupling is of interest for study of nonlinear dynamical systems, as well as a practical tool for controlling and transforming light in photonic devices and improving the sensitivity of photonic sensors.

We would like to highlight the following complementary work examining the phases of output signals during SSB \cite{Pascal2024arXiv}.

%$Acknowledgemens$.

\begin{acknowledgments}
We thank Prof. Vladimir V. Kocharovsky for fruitful discussion and valuable comments.
The work was supported by the Russian Science Foundation, grant No. 20-72-10188-P.
\end{acknowledgments}

\end{document}

\preprint{APS/123-QED}

\title{Supplemental material to\\ Phase-sensitive symmetry breaking in bidirectionally pumped Kerr microresonators}% Force line breaks with \\

\author{Elena A. Anashkina and Alexey V. Andrianov}
\affiliation{A.V. Gaponov-Grekhov Institute of Applied Physics of the Russian Academy of Sciences
Nizhny Novgorod, 603950, Russia \\
The authors contributed equally to this work.}%

% \email{email@gmail.com}


\maketitle
\section{Stationary states of light at different relative pump phases}

\begin{figure*}
\includegraphics[width=2\columnwidth]{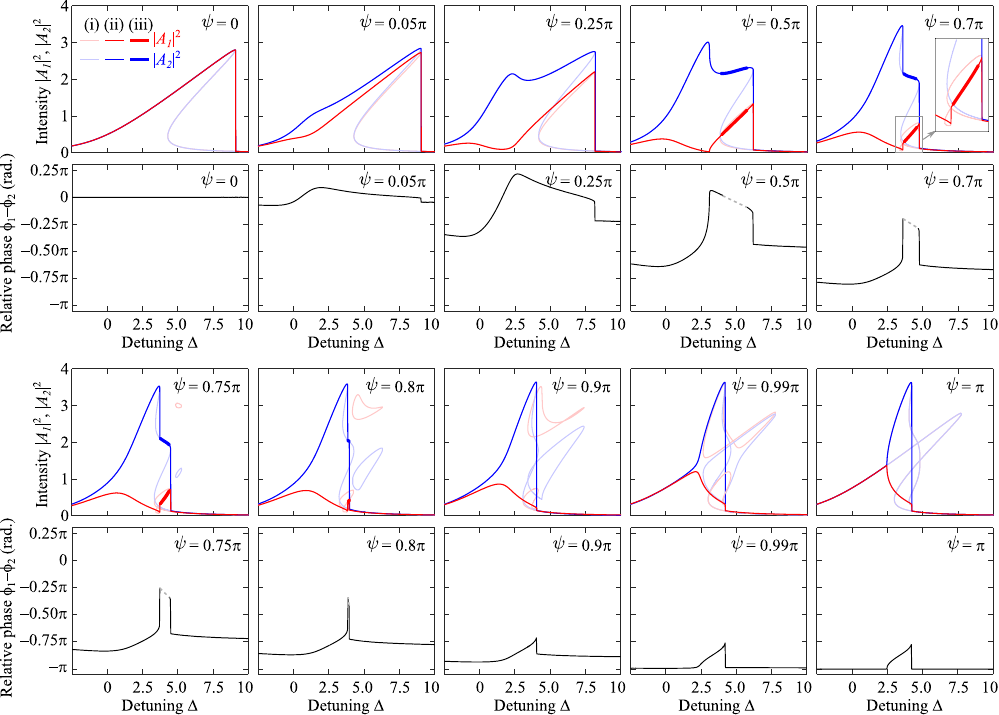}
\caption{\label{fig:L2} Theoretically calculated intensities $|A_1|^2, |A_2|^2$ and difference in phases ($\phi_1-\phi_2$) of counterpropagating modes at $X = 2.8$ and $\beta = 0.65$ and at different relative pump phases $\psi$ indicated in each panel. Light red and light blue lines (i) in the first and third rows demonstrate all stationary solutions. Thin red and blue lines (ii) in the first and third rows show stable CW states numerically found by integrating dynamical system for increasing detuning, while thick lines (iii) correspond to self-oscillation regimes (whithout indication of their amplitudes). Solid black lines in the second and fourth rows show phase difference ($\phi_1-\phi_2$) for realized steady states. Dotted gray lines in the second and fourth rows at $\psi = 0.5\pi, 0.7\pi, 0.75\pi$, and $0.8\pi$ correspond to self-oscillation regimes, where ($\phi_1-\phi_2$) is not constant at slow time (see example for $\psi = 0.5\pi, \Delta = 5.4$ in Fig. S1(e)), so, is not determined by $\Delta$.}
\end{figure*}

To better understand the experimental and numerical results on the formation of deterministic and nondeterministic states of light at different relative pump phases $\psi$ presented in the main text, we theoretically found all  stationary states of the system of 
master equations (1)(a,b), written as a system of nonlinear algebraic equations:
\begin{subequations}
	\label{eq:Eq1}
\begin{eqnarray}
(-1-i\Delta+i|A_1|^2+2i|A_2|^2)A_1+
i\beta A_2 + \sqrt{X}=0,\nonumber\\
(-1-i\Delta+2i|A_1|^2+i|A_2|^2)A_2+
i\beta A_1 + \sqrt{X}e^{i\psi}=0\nonumber.
\end{eqnarray}
\end{subequations} 
We consider the system behavior at $X = 2.8$, $\beta = 0.65$ as for many examples in the main text. The stationary solutions at different phases $\psi$ are shown in Fig.~S1 (see the first and third rows) for $|A_1|^2$ and $|A_2|^2$ by light red and light blue curves, respectively. The maximum number of stationary states is 9 at a relative pump phase near $\pi$ in a narrow range of detunings. We also independently integrated the dynamic system of equations~(1)(a,b) at different detunings for sufficiently long times, as a result of which the system came to a steady state or achieved self-oscillation regimes for each specific $\Delta$.  The corresponding stable steady states of $|A_1|^2$ and $|A_2|^2$ found by numerical integration of the dynamic system are shown by red and blue thin lines, respectively, in Fig.~S1 (the first and third rows). The areas shown by thick lines in the panels for $\psi = 0.5, 0.7, 0.75$, and $0.8$ correspond to self-oscillations (without indication of their amplitudes). Stable stationary solutions found by integrating a dynamic system coincide with analytical solutions (at $d/dt = 0$) on the corresponding branches. Sharp jumps in the red and blue curves correspond to the transition from one stationary branch to another, when the branch ends or the point at which the solution loses stability. At $\psi = 0$, the solutions $|A_1|^2$ and $|A_2|^2$ coincide completely. At $\psi = \pi$, spontaneous symmetry breaking is observed; the pitchfork bifurcation occurs at $\Delta \approx  2.4$; it is randomly determined for which wave the intensity will be higher and for which lower (as discussed in the main text). The calculated relative phases ($\phi_1-\phi_2$) between couterpropagating modes ($A_1 = |A_1|$exp$(i\phi_1)$, $A_2 = |A_2|$exp$(i\phi_2)$) are also plotted in Fig. S1 in the second and fourth rows. At $\psi=0$, completely symmetrical states are realized for which $A_1=A_2$, $\phi_1-\phi_2=0$. At $\psi = \pi$, in the detuning region where the intensities coincide $|A_1|^2=|A_2|^2$, the phases $\phi_1$ and $\phi_2$ differ by $\pi$. When the light states of $|A_1|^2, |A_2|^2$ are asymmetric (i.e. $|A_1|^2 \neq |A_2|^2$), the phase difference ($\phi_1-\phi_2$) changes as shown in Fig. S1.

Let us trace the evolution of stationary solutions with the relative pump phase increasing in the $0 < \psi < \pi$  range, when Eqs.~(1)(a,b) lose their symmetry with respect to $|A_1|^2$, $|A_2|^2$ (Fig.~\ref{fig:L2}).  The powers $|A_1|^2$ and $|A_2|^2$ differ for all $\Delta$ (although  $|A_1|^2$ and $|A_2|^2$ are very close on the lowest branch at large $\Delta$ after passing the resonance, but this section is not interesting). For small $\psi$, the solutions $|A_1|^2$ and $|A_2|^2$ are only slightly deformed relative to the solution at $\psi = 0$, while the power in the second mode is slightly higher than in the first one. As the relative phase increases, the difference in power becomes more significant and depends on $\Delta$. The topology of stationary solutions becomes complex. For example, at $\psi = 0.5\pi$, the solution for $|A_1|^2$ contains a loop, and at $\psi = 0.7\pi$ it has two loops, which are also shown in the inset. For a further phase increase, the solution for $|A_1|^2$ also has two self-intersections. Isolas are born in the third row in Fig.~S1. To understand their origin, let us see the panel for $\psi = \pi$, where the solutions are degenerated. With a small change in phase, the degeneracy is removed, the branches for $|A_1|^2$ and $|A_2|^2$ are separated, isolas with self-intersection are formed, which deform and decrease with a further decrease in phase, and then completely disappear. These isolas have stable sections, which was verified by modelling dynamic equations~(1)(a,b) with initial conditions near the found stationary solutions. However, with smooth frequency scanning from “minus infinity”, these states are not achieved.

Despite a very complex  structure of stationary solutions and system multistability, modelling of a dynamic system shows that with frequency scanning (by increasing $\Delta$), the following scenario for passing the nonlinear resonance with deterministic behavior of $|A_1|^2$ and $|A_2|^2$ is realized (Fig.~S1). The intensity $|A_2|^2$ is higher than the intensity $|A_1|^2$ for $0 < \psi < \pi$; the evolution of $|A_2|^2$ along the topmost branch to its end is implemented. After the topmost branch ends, the solution for $|A_2|^2$ can either jump to the lowest branch or to the middle one. After passing through the section of the middle branch, the deform and decrease the $|A_2|^2$ power drops to the lowest one. The blue resonance curve is deterministically higher than the red one for $0 < \psi < \pi$; the intensity $|A_2|^2$ is higher than the intensity $|A_1|^2$. When replacing the phase $\psi$ with $-\psi$, the red and blue branches change places; therefore, Fig.~S1 shows the range of relative phases only from 0 to $\pi$.

\section{Experimental details}

\begin{figure}
\includegraphics[width=1\columnwidth]{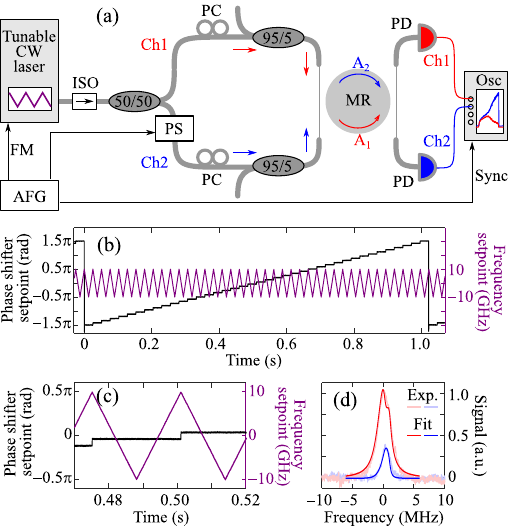}
\caption{\label{fig:L3} (a) Simplified experimental scheme. ISO – Faraday isolator; PS – phase shifter; PC – polarization controller; MR – microresonator; PD – photodiode; Osc – oscilloscope; Ch1 – channel 1; Ch2 – channel 2, AFG – arbitrary function generator; FM – frequency modulation; sync – synchronization. (b,c) Phase shifter setpoint (black curve, left axis) and frequency setpoint of tunable CW laser (purple curve, right axis) as functions of time. Panel (c) demonstrates zoomed in part of panel (b). (d) Almost linear resonance under a unidirectional pump measured in the copropagating direction (light red curve) and in the counterpropagating one (light blue curve) fitted for $\beta = 0.65$, $X = 0.6$.}
\end{figure}

The scheme of our experimental setup with additional details is shown in Fig.~\ref{fig:L3}(a). We used a silica microsphere with a diameter of 120 $\mu$m made of a telecom fiber with the help of a fiber splicer. The calculated zero dispersion wavelength for this microsphere was 1.63 $\mu$m, which allowed us to operate in the region with high normal dispersion and avoid modulation instability. To bidirectionally pump the microsphere, a tunable laser (Santec TSL-770) was used. Its radiation was divided equally into two channels and was coupled into the microsphere in opposite directions through a silica taper. The fiber lengths in the channels were made equal relative to the position of the microsphere on the taper with a precision of 2 mm to ensure minimal relative phase variation over a frequency span of several GHz. A piezoelectric phase shifter with a characteristic response time of 0.1 ms was placed in channel 2. To measure intraresonator fields we used a second taper in a ``drop-port" configuration and two identical photodetectors attached to an oscilloscope (Tektronix MSO-64 6-BW-8000). 
The optical spectrum analyzer attached to one of the ports was used to monitor that nonlinear optical generation of new frequencies  should not occur in the microresonator, either near the pump wavelength, or in the Raman gain band.

The laser frequency and phase shift were controlled using an arbitrary function generator. The phase difference introduced by the phase shifter had the staircase profile covering the range from $-1.5\pi$ to $1.5 \pi$ in 40 small steps (see Fig.~\ref{fig:L3}(b)). The commanded laser frequency varied in the range from about $-10$ to 10 GHz in a sawtooth profile synchronized with the phase jumps as shown in Fig.~\ref{fig:L3}(c). The mean laser frequency was adjusted so that it was near a certain resonance. With these settings we collected 40 frames per slow phase scan, each frame containing the microresonator response to the pumps with almost constant phase shift between opposite directions, while the frequency was swept up and down. Although the absolute phase between the pumps arriving at the microresonator could not be measured directly and drifted slowly, it was stable enough over the phase scan period, and the absolute phase could be identified later by inspecting certain features of the recorded signals.

We selected a certain resonance around 1483.3 nm and worked only at this resonance in all experiments. Figure~\ref{fig:L3}(d) shows this resonance measured through the drop port under unidirectional pump with reduced power in an almost linear regime. In this case, the loaded Q-factor was about $10^8$, and the intermode Rayleigh scattering coefficient ensuring the best agreement between the measurement and the simulation was $\beta = 0.65$. Both tapers were adjusted to operate in undercoupled regime so they did not make a significant impact on the Q-factor.

%\maketitle
\section{Self-oscillation regimes}

\begin{figure}
\includegraphics[width=0.92\columnwidth]{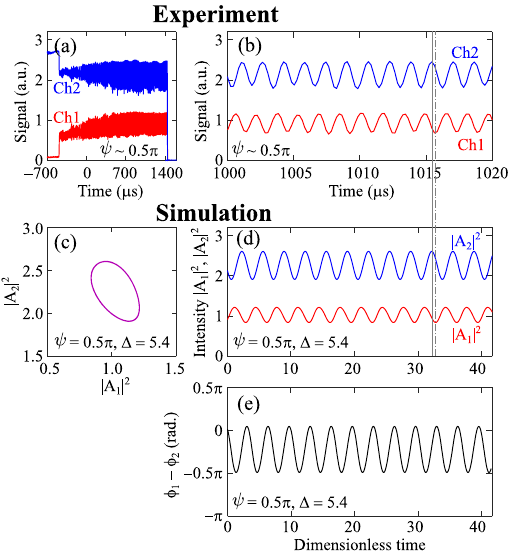} 
\caption{\label{fig:L6}  (a,b) Experimental signals measured by oscilloscope in channel 1 (red curves) and channel 2 (blue curves) demonstrating self-oscillations for relative phase $\psi \sim 0.5\pi$. Experiment was performed according to the scheme shown in Fig. 3(a) in the main text. (c) The simulated phase space orbit of intensities $|A_1|^2$, $|A_2|^2$ for self-oscillations and (d) the corresponding oscillograms at $\psi ~ 0.5\pi$, $\Delta = 5.4$, and $X = 2.8$.  (e) Relative phase $\phi_1-\phi_2$ between counterpropagating modes as a function of slow time. The solid vertical line through (b,d) shows signal maximum in channel 2 ($|A_2|^2$), while the dash-dotted line shows signal minimum in channel 1 ($|A_1|^2$).}
\end{figure}

Regions of the parameters in which self-oscillations are observed were found theoretically and experimentally. Here we will pay attention to this mode.
We recorded in detail the oscillogram of self-oscillations near $\psi \approx 0.5\pi$ by significantly slowing down the frequency scan when reaching the corresponding region by modifying the fast scanning frequency function of the lasers. The experimental oscillogram is shown in Fig.~S3(a). The section from 0 to 1400 $\mu$s contains about 900 oscillation periods. A zoomed-in section is shown in Fig.~S3(b). The oscillations are not exactly out-of-phase; the maximum signal in channel 1 and the minimum one in channel 2 are achieved at different times, as demonstrated by the vertical lines. The experimental results are in a very good agreement with the simulation results presented in Fig.~S3(c,d). The oscillograms are very similar, and the oscillation phases coincide (Fig.~S3(b,d) vertical lines). The simulated dependence of the phase trajectory is shown in Fig.~S3(c). The self-oscillations are not harmonic and the phase difference ($\phi_1-\phi_2$) depends on slow time (Fig. S3(e)).